\documentclass[12pt]{article}
\begin{document}
\title{Stability of  a Nonequilibrium Interface in a Driven
Phase Segregating System}
\author{Claude A. Laberge and Sven Sandow \\[0.5cm]
   {\it \small Department of Physics 
  and Center for Stochastic Processes in Science and Engineering}\\
 {\it \small Virginia Polytechnic Institute and State University}\\
  {\it \small Blacksburg, VA 24061-0435, USA }}

\vspace{2cm}
\maketitle
\vspace{3cm}

\begin{abstract}
We investigate the dynamics of a nonequilibrium interface between coexisting
phases in a system described by a Cahn-Hilliard equation with
an additional driving term. By means of a matched
asymptotic expansion we derive equations for the interface motion. A linear
stability analysis of these equations results in a condition for the
stability of a flat interface. We find that the stability properties
of a flat interface depend on the structure of the driving term
in the original equation.

\end{abstract}

\vspace{31mm}
\rule{7cm}{0.2mm}
\begin{flushleft}
\parbox[t]{3.5cm}{\bf Key words:}
\parbox[t]{12.5cm}{phase segregation, nonequilibrium interface, instability} 
		   \\[1mm]
\parbox[t]{3.5cm}{\bf PACS numbers:} 
\parbox[t]{12.5cm}{64.75.+g, 64.70.Ja, 68.10.-m, 47.54.+r}
\end{flushleft}
\normalsize
\thispagestyle{empty}
\mbox{}
\pagestyle{plain}

\newpage

\section{Introduction}
Off-equilibrium systems composed of regions with different phases can exhibit a variety of patterns,
such as fingers aligned in a certain direction.
A familiar example is spinodal decomposition of binary mixtures
 (for a review see \cite{gss}, \cite{langer}). When quenched
below its critical temperature an initially homogeneous system develops
domains of the new equilibrium phases. After some time these domains
are well defined, i.e. they are separated by sharp interfaces. 
The system further approaches equilibrium by means of the motion of these 
interfaces, the initial pattern evolves in the course of time.
 Another example is the dynamics of driven diffusive systems 
(for a review see Ref. \cite{sz}). Interacting particles undergoing 
biased diffusion tend to form clusters under certain conditions. Like
for the first example there are well defined interfaces, which
undergo some dynamics.  In all of these phase segregating systems
patterns are selected according to their stability. 
For example the instability of a flat interface causes the
growth of fingers  out of an initially flat interface. For this reason
one would like to understand the stability properties of 
nonequilibrium interfaces.

On a course grained level these phase segregating systems are described
by means of an order parameter, such as a particle density. Typically, the 
dynamics of this order parameter can be  modeled by means of some
nonlinear differential equation. One of the simplest such equation
is the Cahn-Hilliard equation (CH equation), which was introduced in the
context of
binary mixtures \cite{ch}. The properties of this equation
are rather well understood. In particular, the dynamics of the domain 
boundaries obey a set of linear differential equations, which have been
derived by Pego \cite{pego} using of a matched asymptotic expansion.
An instability of the Mullins-Sekerka type has been observed.

Certain modification of the CH-equation
were introduced to take into account the effect of
external fields \cite{koj}-\cite{ymhj}, \cite{eb}
(see  Refs. \cite{gl}, \cite{allz}, \cite{sz} for modified CH equations
in the context of driven diffusive systems). One of the questions
one would like to answer is the following: How does an external field
change the stability properties of a flat interface? This question was addressed
by  Yeung et. al. \cite{ymhj} for a certain type of driving term. This driving term
has the form of  an additional current which is proportional to
the field and a conductivity. The conductivity was assumed to be   a concave second order function of the order parameter with the same symmetry as the chemical potential. Yeung et. al. \cite{ymhj} found that the field  modifies the
interface dynamics. An interface perpendicular to the external field is stable for one direction of
the field and unstable for the other direction.

In this paper we generalize the work of Yeung et. al. \cite{ymhj}. We assume
a more general structure of the driving term, i.e. a more general conductivity.
Our approach is the same as the one from Ref. \cite{ymhj} (see also Ref. \cite{pego}). Assuming that
the driving field is sufficiently small we perform a matched asymptotic
 expansion of a modified Cahn-Hilliard equation with an arbitrary conductivity. The resulting set of linear equations describes the motion of an interface
between coexisting phases. With the help of these equation we analyze the
stability of a flat interface. 
We find a general stability criterion, 
which takes a very simple form if we assume that the conductivity is
a third order polynomial in the order parameter. For such a conductivity
we show that an interface perpendicular to the external field can be unstable
for either direction of the field, and that an interface parallel to the external field is always stable.

The paper is organized as follows. In Section 2 we define the model. A matched
asymptotic expansion is done in Section 3. Equations for the motion
of an interface are derived. In Section 4 we perform a linear stability
analysis for a flat interface resulting in a condition for the stability
of such an interface. This condition is subsequently shown to take a simple form
 for the case of a  conductivity that is a third order polynomial in
the order parameter.  Section 5 summarizes  the results.

\section{The model}

We consider a two-dimensional system described by a 
conserved order parameter $u({\bf R},T)$
where ${\bf R}$ denotes a position in space and $T$ the time.
The dynamic is assumed to be defined by the following modified
Cahn-Hilliard equation \cite{ppd}-\cite{bray}, \cite{ymhj}, \cite{eb}, \cite{allz}:
\begin{eqnarray}
\label{continuity}
\frac{\partial u}{\partial T}&=& -\nabla \cdot {\bf J} \\
\label{current}
{\bf J} &=&- \nabla \mu + {\bf E}~\sigma(u) \\
\label{mu}
\mu &=&\lambda u (u-\frac{u_m}{2})(u-u_m) 
- \xi^2 \nabla^2 u ~.
\end{eqnarray}
In the above equations, ${\bf J}$ is the 
current of the order parameter,  $\mu$ is the chemical potential
of the free system, 
${\bf E}$ is an external field (e.g. an electric field), and 
$\sigma(u)$ is the conductivity.  One may think of the chemical
potential as related  by means of
$\mu=\delta F/ \delta u$ to
 a free energy functional 
$F[u]=\int d{\bf R}\;\{\; f(u({\bf R}))+  
\xi^2 |\nabla u({\bf R})|^2 /2 \;\}\;$, where $f(u)$ is a bulk free energy density
the derivative of which is
$f'(u)=\lambda u (u-\frac{u_m}{2})(u-u_m)\;$.
 The current has two additive 
contributions: One that minimizes the free energy and another one 
defining the transport caused by the field.  The first one is specified 
by the structure of the chemical potential.  Assuming  
$\lambda>0$ and $u_m>0$ 
Eq. (\ref{mu}) defines a chemical potential such that the free energy has 
minima in 
regions with $u=0$ or with $u=u_m>0$. The term proportional to 
$\xi^2$ results
in an increase of energy whenever there is a gradient in $u$, 
i.e., $\xi^2$ stands for a surface tension. 
The current caused by the field is  ${\bf E}~\sigma(u)\;$,  it depends
on the function $\sigma(u)\;$. The standard conductivity used
in the literature \cite{ppd}-\cite{bray}, \cite{ymhj}, \cite{eb}, \cite{allz} is a concave second order function of $u\;$, which 
has the same symmetry as the chemical potential, i.e.
is symmetric around $u_m/2$. Here, we do not restrict to this type of
function, but rather want to see what kind of behavior can be observed
for a more general conductivity.

Equations of type (\ref{continuity})-(\ref{mu}) are usually motivated on a purely
phenomenological (see e.g. \cite{lhs,bray,ymhj,eb,allz}) 
or mean field (see e.g. \cite{ppd}) level.
The standard type of argument results in  a mobility as a factor
to the total current, i.e. a current of the type  
$\sigma(u)[- \nabla \mu + {\bf E}] \;$. Assuming furthermore that
the variation of $\sigma(u)$ is small enough $\sigma(u) \nabla \mu$
is then replaced by $ \nabla \mu\;$. We follow the same logic here.
It is also worthwhile mentioning that a rigorous derivation of a more general type of equation  has been done 
for diffusive system with long-range interaction and Kawasaki dynamics \cite{gl}.

Since it is more convenient to work with a dimensionless equation we rescale
Eqs. (\ref{continuity})-(\ref{mu}). Simultaneously we add a constant
term to the conductivity and apply a Galilean 
transformation such that  the new  current vanishes at both 
minima of the chemical potential. The adding of a constant does not change the equation
for $u({\bf R},T)\;$, since only the gradient of $\sigma$ enters into this 
equation. The Galilean  transformation is nothing but a change of
the reference frame.  
We define a length $L_0$ and  a time $T_0$ obeying
\begin{equation}
\label{tl}
1=L_0^{-2} T_0 \lambda u_m^2
\end{equation}
as well as a velocity 
\begin{eqnarray}
\label{vg}
V_g=|{\bf E}|~u_m^{-1}~\sigma(u_m)~.
\end{eqnarray} 
Now we define dimensionless quantities as:
\begin{eqnarray}
\label{trafo_r}
{\bf r} &=& L_0^{-1}~{\bf R} - L_0^{-1} V_g~T~{\bf e} \\
\label{trafo_t}
t &=& T_0^{-1}~T \\
\label{trafo_rho}
\rho &=& u_m^{-1}~u ~.
\end{eqnarray}
where
\begin{equation}
\label{unitvector}
{\bf e} = \frac{{\bf E}}{|{\bf E}|}
\end{equation}
is the unit vector in the direction of the field. Inserting the 
above definitions into Eqs. (\ref{continuity})-(\ref{mu}) yields
the following dimensionless equation
\begin{eqnarray}
\label{continuity1}
\frac{\partial \rho}{\partial t}&=& -\nabla \cdot {\bf j} \\
\label{current1}
{\bf j} &=& -\nabla \nu + {\bf e}~ \beta(\rho) \\
\label{mu1}
\nu &=&\rho (\rho -1)(\rho-\frac{1}{2}) - \epsilon^2 \nabla^2 \rho 
\end{eqnarray}
with
\begin{equation}
\label{epsilon}
\epsilon^2 = L_0^{-2} \xi^2 \lambda^{-1} u_m^{-2} 
\end{equation}
and
\begin{equation}
\label{sigma1}
\beta(\rho) = \epsilon^{-1} |{\bf E}| \xi \lambda^{-3/2} 
u_m^{-4}~
\{\;\sigma(\rho u_m)-\sigma(0)\; - \;\rho [\sigma(u_m)-\sigma(0)]\;\}~.
\end{equation}
In Eqs. (\ref{continuity1})-(\ref{mu1}) the operator $\nabla$ stands for
derivative with respect to the new coordinate ${\bf r}\;$.
The conductivity $\beta(\rho)$ in the new frame has the property
\begin{eqnarray}
\label{v123}
\beta(1)=\beta(0)=0
\end{eqnarray}
i.e., it vanishes at both minima of $\nu\;$. However it is not 
necessarily positive in the whole interval $(0,1)\;$. In fact, later on 
we will identify a regime where $\beta(\rho)<0$
for some part of $(0,1)\;$. For that reason it would be hard to give a direct
physical meaning to Eqs. (\ref{continuity1})-(\ref{sigma1}) in terms of 
the rescaled variables. 

The velocity $V_g$ given by Eq. (\ref{vg}) is positive by definition.  
Consequently the new frame of reference moves in the  direction of the 
field with respect to the original one.

\section{Dynamics of an interface}

The model defined in the previous section describes phase segregation, 
i.e., the order parameter $\rho({\bf r},t)$ evolves such that there are
regions in space where $\rho({\bf r},t)$  approximately takes the values 
for which the chemical potential is minimal. 
Since there are two minima of the chemical potential, there are
two types of regions, which are separated by transition layers.  The transition
layer between the different regions, which is seen as a line on a large
enough scale,  evolves during time. We are going to give an 
analytical description of its motion for the case where
\begin{eqnarray}
\label{cond eps}
\epsilon &\ll& 1\\
\label{cond sigma}
\mbox{  and for }\;0\le \rho \le 1\;:
\;\;\;\beta(\rho)&=& O(\epsilon^0)\;\;\mbox {and }\;\;
\beta'(\rho)= O(\epsilon^0)
\end{eqnarray}
in the rescaled equations (\ref{continuity1})-(\ref{sigma1}). In terms of 
the original parameters these conditions read
\begin{eqnarray}
\label{cond 1}
L_0^{-1} \xi \lambda^{-\frac{1}{2}} u_m^{-1}& \ll& 1 \\
\label{cond 2}
\mbox{  and  }\;\;\;|{\bf E}|\;\xi\;\lambda^{-3/2} 
u_m^{-4}\;\sigma_{max}&\ll & 1
\;\;\end{eqnarray}
where $\sigma_{max}$ denotes the largest values $\sigma(u)$ 
and its derivative $\sigma'(u)$ take in $[0,1]\;$.
The first of these conditions can be fulfilled for any choice of 
the parameters simply by choosing a large enough length scale $L_0\;$. 
However, the second condition poses a restriction on the parameters, it can 
be understood as a small field condition.

Starting with the rescaled equations (\ref{continuity1})-(\ref{mu1})  
we apply a matched asymptotic expansion. The basic idea  is 
to expand the order parameter and its evolution  equation in powers of 
the small parameter $\epsilon$. We closely follow the line of Refs. 
\cite{pego} and \cite{ymhj}. Since we want to describe the motion of 
a single interface, we consider an initial configuration with two 
semi-infinite regions $\Omega^+$ and $\Omega^-$ 
and assume that the order parameter is equal to $1$ ($0$) in $\Omega^+$ 
($\Omega^-$) up to corrections of order $\epsilon\;$. The values $1$ 
and $0$ are the minima of the chemical potential. The regions $\Omega^+$ 
and $\Omega^-$ are separated by a transition layer.
We assume this layer  to 
have a width of order $\epsilon\;$ like it in the field free case \cite{pego}, \cite{ymhj}. The characteristic time of its motion is
\begin{equation}
\label{tau}
\tau= \epsilon~t~=~\epsilon~T_0^{-1}~T ~~
\;.\end{equation}
In order to have a clear definition on any scale
we define the interface $\Gamma$ as the set of points 
where $\rho=\frac{1}{2}\;$. (Our results are not sensitive to the choice of the
value $1/2\;$.) 

We are going to expand our equations on the time scale $\tau$
first in the bulk of $\Omega^+$ 
and $\Omega^-$ and then in the transition layer, i.e. near
the interface $\Gamma\;$. 

\subsection{Equations far from the interface}

Far from the interface $\Gamma$ we define the new field 
\begin{equation}
\label{rhopm}
\rho^{\pm}({\bf r},\tau)=\rho({\bf r},\epsilon^{-1}t)
\end{equation}
and similarly $\nu^{\pm}({\bf r},\tau)$ and $\beta^{\pm}({\bf r},\tau)$
where the superscript denotes the region we are considering.
Eqs. (\ref{continuity1}) and (\ref{current1}) result in the following
equations on the time scale $\tau$
\begin{eqnarray}
\label{continuity2}
\epsilon~\partial_{\tau} \rho^{\pm}({\bf r},\tau)&=&
\nabla^2 \nu^{\pm}({\bf r},\tau) -
{\bf e}\cdot\nabla \beta^{\pm}({\bf r},\tau)
\end{eqnarray}
where $\nu^{\pm}$ and $\beta^{\pm}$ are defined  by Eqs. (\ref{mu1}) and
(\ref{sigma1}), respectively. 
Next we expand $\rho^{\pm}({\bf r},\tau)$ in powers of $\epsilon$ as 
\begin{equation}
\label{expand rho}
\rho^{\pm}({\bf r},\tau) = \rho_0^{\pm}+
\epsilon~\rho_1^{\pm}({\bf r},\tau)+O(\epsilon^2)
\end{equation}
where 
\begin{equation}
\label{rhopm0}
\rho_0^+=1~~\mbox{and}~~\rho_0^-=0
\end{equation}
according to our previous assumptions. Inserting above expansion into 
Eq. (\ref{continuity2}), (\ref{mu1}) and (\ref{sigma1}) results in an 
equation for $\rho_1^{\pm}({\bf r},\tau)$ where the terms of zeroth order in 
$\epsilon$ cancel each other. Comparing terms of first order in $\epsilon$ 
yields the following relation
\begin{equation}
\label{rho_1}
[~\nabla^2 - 2 B^{\pm}~{\bf e}\cdot\nabla~]~\rho_1^{\pm}({\bf r},\tau)=0
\end{equation}
with 
\begin{equation}
\label{B+-}
B^+=\beta'(1)~\mbox{and}~B^-=\beta'(0)~~.
\end{equation}

\subsection{Equations near the interface}

We  are going to expand Eqs. (\ref{continuity1})-(\ref{mu1}) in the 
transition layer where the chemical potential is not close to its minima.
Let us denote by ${\bf r}_{\Gamma}$ a point there, i.e. a point with a 
distance of order $\epsilon$ from $\Gamma$, and introduce the quantity 
$\Phi({\bf r}_{\Gamma},\tau)$ as the signed distance between point 
${\bf r}_{\Gamma}$ and the interface $\Gamma$ at the time $t=\epsilon^{-1} 
\tau$. The sign is chosen such that $\Phi>0$ in $\Omega^+$ and  
$\Phi<0$ in $\Omega^-$. Furthermore we define
\begin{eqnarray} 
\label{m}
{\bf m}({\bf r}_{\Gamma},\tau)&=&\nabla \Phi({\bf r}_{\Gamma},\tau)\\
\label{kappa}
\kappa({\bf r}_{\Gamma},\tau)&=& \nabla^2 \Phi({\bf r}_{\Gamma},\tau)\\
\label{v}
V({\bf r}_{\Gamma},\tau)&=&\partial_{\tau} \Phi({\bf r}_{\Gamma},\tau)
~~.\end{eqnarray}
As can be seen from these definitions, if ${\bf r}_{\Gamma}$ lies on 
$\Gamma$ the vector ${\bf m}$ is the unit normal of the interface at  
${\bf r}_{\Gamma}$, $\kappa$ is the curvature there and $V$ is the normal 
velocity of $\Gamma$ on the time scale $\tau$. Signs are such that 
the vector ${\bf m}$  points into $\Omega^+$, the curvature $\kappa$
is positive if the center of curvature lies in $\Omega^-$ and $V$ is 
positive if the interface moves towards $\Omega^-$
 
Since the transition layer has a width of order $\epsilon$ it is 
convenient to introduce a variable
\begin{equation}
\label{z}
z=\epsilon^{-1} \Phi({\bf r}_{\Gamma},\tau)
\end{equation}
and for an arbitrary 
field $f({\bf r}_{\Gamma},t)$ a field  $\tilde {f}(z,{\bf r}_{\Gamma},\tau)$ by:
\begin{equation}
\label{trafo_tilde}
f({\bf r}_{\Gamma},t)=\tilde {f}(\epsilon^{-1} 
\Phi({\bf r}_{\Gamma}),{\bf r}_{\Gamma}, \epsilon~t)
\end{equation}
The derivatives occurring in our equations are transformed as follows:
\begin{eqnarray}
\label{gradient}
\nabla f&=&[~\nabla_r +\epsilon^{-1} {\bf m} \partial_z~]~\tilde{f}\\
\label{gradient^2}
\nabla^2 f&=&[~\nabla_r^2 +\epsilon^{-1} \kappa \partial_z+
\epsilon^{-2}\partial_z^2~]~\tilde{f}\\
\label{dbydt}
\partial_t f&=&[~\epsilon~\partial_{\tau} + V~\partial_z~]~\tilde{f}
\end{eqnarray}
where $\nabla_r$ is the gradient acting on ${\bf r}_{\Gamma}$ only, and 
arguments were dropped for convenience. Assuming furthermore that all 
relevant fields near $\Gamma$ depend
only on their relative position with respect 
 to $\Gamma$, i.e., that
$\tilde {f}(z,{\bf r}_{\Gamma}+\delta \;{\bf m}({\bf r}_{\Gamma},\tau),\tau)=
\tilde {f}(z,{\bf r}_{\Gamma},\tau)$  for any small number $\delta$, we find
\begin{equation}
\label{null}
\nabla \Phi({\bf r}_{\Gamma},\tau) \cdot  
\nabla_r \tilde {f}(z,{\bf r}_{\Gamma},\tau)=0 ~~.
\end{equation}

With the above relations we are ready to transform and expand in powers 
of $\epsilon$ as:
\begin{eqnarray}
\label{expand rho1}
\rho&=~\tilde{\rho}~=&\tilde{\rho}_0+
\epsilon~\tilde{\rho}_1+\epsilon^2 \tilde{\rho}_2+O(\epsilon^3)\\
\label{expand mu1}
\nu&=~\tilde{\nu}~=&\tilde{\nu}_0+
\epsilon~\tilde{\nu}_1+\epsilon^2 \tilde{\nu}_2+O(\epsilon^3)\\
\label{expand sigma1}
\beta&=~\tilde{\beta}~=&\tilde{\beta}_0+
\epsilon~\tilde{\beta}_1+\epsilon^2 \tilde{\beta}_2+O(\epsilon^3)
\end{eqnarray}
where the arguments, which we have dropped for convenience, 
have to be taken according to Eq. (\ref{trafo_tilde}). 
The expansions of $\nu$ and $\sigma$ are
related to the one of $\rho$ by means of Eqs. (\ref{mu1}) and (\ref{sigma1}). 
These relations can be casted in an explicit form using Eqs. (\ref{gradient}),
(\ref{gradient^2}), (\ref{dbydt}) and (\ref{null}). The following ones we
are going to use later:
\begin{eqnarray}
\label{mu0_new}
\tilde{\nu}_0&=&\tilde{\rho}_0(\tilde{\rho}_0-1)(\tilde{\rho}_0-\frac{1}{2})~-~
\partial_z^2 \tilde{\rho}_0\\
\label{mu1_new}
\tilde{\nu}_1&=&[3 \tilde{\rho}_0(\tilde{\rho}_0-1)+\frac{1}{2}]~\tilde{\rho}_1
- \kappa \partial_z \tilde{\rho}_0~-~\partial_z^2 \tilde{\rho}_1\\
\label{mu2_new}
\tilde{\nu}_2&=&[3 \tilde{\rho}_0(\tilde{\rho}_0-1)+\frac{1}{2}]~\tilde{\rho}_2
+\frac{3}{2}(2 \tilde{\rho}_0-1) \tilde{\rho}_1^2 
-\nabla_r^2 \tilde{\rho}_0 - \kappa \partial_z \tilde{\rho}_1 -
\partial_z^2 \tilde{\rho}_2\\
\label{sigma01}
\tilde{\beta}_0&=&\beta( \tilde{\rho}_0)\\
\label{sigma11}
\tilde{\beta}_1&=&\tilde{\rho}_1~\beta'( \tilde{\rho}_0)~~.
\end{eqnarray}
The last two of these relations hold under the condition (\ref{cond sigma}) 
only. Inserting now expressions (\ref{gradient}), (\ref{gradient^2}) 
and (\ref{dbydt}) as well as expansions (\ref{expand rho1}), 
(\ref{expand mu1}) and (\ref{expand sigma1}) into our evolution 
equation (\ref{continuity1}), (\ref{current1}) and comparing terms of 
order $\epsilon^{-2}~,\epsilon^{-1}~,\epsilon^{0}$, respectively, 
results in the following set of equations:
\begin{eqnarray}
\label{new1}
0&=&\partial_z^2 \tilde{\nu}_0\\
\label{new2}
0&=&\kappa~\partial_z \tilde{\nu}_0 + \partial_z^2 \tilde{\nu}_1 -
{\bf e}\cdot{\bf m}~\partial_z \tilde{\beta}_0\\
\label{new3}
V \partial_z \tilde{\rho}_0&=& \nabla_r^2 \tilde{\nu}_0 +
\kappa~\partial_z \tilde{\nu}_1 + \partial_z^2 \tilde{\nu}_2 -
{\bf e}\cdot\nabla_r \tilde{\beta}_0 - 
{\bf e}\cdot{\bf m}~\partial_z \tilde{\beta}_1~~.
\end{eqnarray}
The dynamics of the order parameter near the interface is described by 
the above
equations combined with the following boundary conditions: 
The solution of above equations should
match $\rho^{\pm}({\bf r}_{\Gamma},\tau)$ in $\Omega^{\pm}$, i.e., outside
the transition layer.
We demand 
\begin{equation}
\label{limit}
\lim_{z \rightarrow \pm \infty} \tilde{\rho}(z,{\bf r}_{\Gamma},\tau)~=~
\rho^{\pm}({\bf r}_{\Gamma},\tau)~~.
\end{equation}
Using 
$\lim_{z \rightarrow \pm \infty} \tilde{\rho}(z,{\bf r}_{\Gamma},\tau)=
\lim_{z \rightarrow \pm \infty}
 {\rho}({\bf r}_{\Gamma}+\epsilon~z~{\bf m},\epsilon^{-1}\tau)$
and expanding above
condition in powers of $\epsilon$ yields
\begin{eqnarray}
\label{limit0}
\lim_{z \rightarrow \pm \infty} \tilde{\rho}_0(z,{\bf r}_{\Gamma},\tau)&=&
\rho^{\pm}_0\\
\label{limit1}
\lim_{z \rightarrow \pm \infty} \tilde{\rho}_1(z,{\bf r}_{\Gamma},\tau)&=&
\rho^{\pm}_1({\bf r}_{\Gamma},\tau)\\
\label{limit2}
\lim_{z \rightarrow \pm \infty} \tilde{\rho}_2(z,{\bf r}_{\Gamma},\tau)&=&
\rho^{\pm}_2({\bf r}_{\Gamma},\tau)~+~
\lim_{z \rightarrow \pm \infty}~[~z~{\bf m}\cdot 
\nabla \rho^{\pm}_1({\bf r}_{\Gamma},\tau)~]
\;.\end{eqnarray}
Here we used Eqs. (\ref{rhopm}), (\ref{expand rho}) and (\ref{rhopm0}).

One can now solve Eqs. (\ref{new1}), (\ref{mu0_new}) with the 
boundary condition  (\ref{limit0}). The result is
\begin{eqnarray}
\label{rho0_sol}
\tilde{\rho}_0(z,{\bf r}_{\Gamma},\tau)&=&
\bigl[~1+~ e^{-z/\sqrt{2}}~\bigr]^{-1}\\
\label{mu0_sol}
\tilde{\nu}_0(z,{\bf r}_{\Gamma},\tau)&=&0
~~.\end{eqnarray}
Let us now define  the following integrals for later use:
\begin{eqnarray}
\label{S}
S&=&\int^{\infty}_{-\infty} [\partial_z \tilde{\rho}_0(z)]^2 dz 
~=~\frac{\sqrt{2}}{12}\\
\label{I}
I_n&=&\int^{\infty}_{-\infty} [\tilde{\rho}_0(z)]^n~
\beta(\tilde{\rho}_0(z))~ dz 
~=~\sqrt{2}~\int_0^1\frac{\rho^{n-1}}{1-\rho}~ \beta(\rho)~ d\rho
~.\end{eqnarray}
The first integral was computed and the second one
was simplified using Eq. (\ref{rho0_sol}).
Next we insert $\tilde{\nu}_0=0$ into Eq. (\ref{new2}) and integrate over $z\;$.
We fix the integration constants by means of the limits 
$z \rightarrow \pm \infty$. In order to compute these limits we
expand 
$\lim_{z \rightarrow \pm \infty} \tilde{\nu}(z,{\bf r}_{\Gamma},\tau)=
\lim_{z \rightarrow \pm \infty}
 {\nu}({\bf r}_{\Gamma}+\epsilon~z~{\bf m},\epsilon^{-1}\tau)$ in powers of $\epsilon$
and use the fact that $\beta(0)=\beta(1)=0\;$. We obtain:
\begin{eqnarray}
\label{dzmu1}
\partial_z \tilde{\nu}_1= {\bf e}\cdot{\bf m} ~\beta_0
\;.\end{eqnarray}
Integrating once more over $z$ and taking again 
the limits $z \rightarrow \pm \infty$ yields a relation between 
$\rho_1^+$ and $\rho_1^-$.
Another relation is obtained by multiplying Eq. (\ref{new2}) by
$\partial_z\tilde{\rho}_0$ and integrating over $z$ from 
$-\infty$ to $+\infty$.
Combining those two  yields
\begin{eqnarray}
\label{u1plus}
\rho_1^+ ({\bf r}_{\Gamma},\tau)&=&
~-~2S~\kappa({\bf r}_{\Gamma},\tau)\;+\;
2 I_1~{\bf e}\cdot{\bf m}({\bf r}_{\Gamma},\tau) \\ 
\label{u1minus}
\rho_1^- ({\bf r}_{\Gamma},\tau)&=&
~-~2S~\kappa({\bf r}_{\Gamma},\tau)\;+\;
2 (I_1-I_0)~{\bf e}\cdot{\bf m}({\bf r}_{\Gamma},\tau)   
\end{eqnarray}
where $I_0$ and $I_1$ are the integrals defined by Eq. (\ref{I}).

In a last step we integrate Eq. (\ref{new3}) and take the limits 
$z \rightarrow \pm \infty$. As a result we obtain an expression for
the interface velocity
\begin{eqnarray}
V({\bf r}_{\Gamma},\tau)&=&
\frac{1}{2}~{\bf m}({\bf r}_{\Gamma},\tau)~
\left.[~ \nabla \rho_1^+ ({\bf r},\tau)
-\nabla \rho_1^- ({\bf r},\tau)~] \right|_{ r=r_{\Gamma}} \nonumber\\
&&+[~I_0+2(B^+-B^-)S~] 
~{\bf e}\cdot{\bf m}({\bf r}_{\Gamma},\tau)~\kappa({\bf r}_{\Gamma},\tau)\nonumber\\
&&-~2~[~(B^+-B^-)I_1+B^-I_0~]
~[{\bf e}\cdot{\bf m}({\bf r}_{\Gamma},\tau)]^2
\label{vsol}
~~\end{eqnarray}
where $B^{\pm}$ 
is defined by Eq. (\ref{B+-}), 
$S$  by Eq. (\ref{S}) and $I_0\;, \;I_1$ 
by Eq. (\ref{I}).

Above equations provide a macroscopic description of the interface dynamics. 
Macroscopic means that the space resolution is of order $\epsilon^0$ so that 
the transition layer can be identified with interface $\Gamma$, i.e., 
any point ${\bf r}_{\Gamma}$ can be considered as lying on $\Gamma$. 
Suppose that at a time $t=\epsilon^{-1} \tau$ there is an interface 
$\Gamma$ with a unit normal ${\bf m}({\bf r}_{\Gamma},\tau)$ and a curvature 
$\kappa({\bf r}_{\Gamma},\tau)$ at its points ${\bf r}_{\Gamma}$. Away from
the interface the order parameter is given by
$\rho^{\pm}({\bf r}_{\Gamma},\tau)=\rho_0^{\pm} +\epsilon \rho_1^{\pm} 
({\bf r}_{\Gamma},\tau)$
where $\rho^+_0=1~,~\rho^-_0=0$ and $\rho_1^{\pm} ({\bf r}_{\Gamma},\tau)$
is given by the solution of Eq. (\ref{rho_1}) with boundary conditions 
on the interface defined
by Eqs. (\ref{u1plus}), (\ref{u1minus}) in terms of 
${\bf m}({\bf r}_{\Gamma},\tau)$ and  $\kappa({\bf r}_{\Gamma},\tau)$. This 
 solution in its turn determines the normal velocity of each interface-point
by means of Eq. (\ref{vsol}). 
\section{Linear Stability Analysis}

As derived in the previous section, 
on the scale $\tau= \epsilon^{-1}~t
~=~\epsilon^{-1}~T_0^{-1}~T$ 
the motion of an interface at ${\bf r} = {\bf r}_{\Gamma}$
separating the two regions $\Omega^+$ (with a particle density
$\rho^+({\bf r},\tau) = 1 + \epsilon \rho_1^+({\bf r},\tau) + O(\epsilon^2)$) 
and $\Omega^-$ (with 
$\rho^-({\bf r},\tau) = 0 + \epsilon \rho_1^-({\bf r},\tau)+ O(\epsilon^2)$ ) 
is given the normal interfacial velocity $V$ of the interface
\begin{eqnarray}
V({\bf r}_{\Gamma},\tau)&=&
\frac{1}{2}~{\bf m}({\bf r}_{\Gamma},\tau)~
\left.[~ \nabla \rho_1^+ ({\bf r},\tau)
-\nabla \rho_1^- ({\bf r},\tau)~] \right|_{ r=r_{\Gamma}} \nonumber\\
&&+[~I_0+2(B^+-B^-)S~] 
~{\bf e}\cdot{\bf m}({\bf r}_{\Gamma},\tau)~\kappa({\bf r}_{\Gamma},\tau)\nonumber\\
&&-~2~[~(B^+-B^-)I_1+B^-I_0~]
~[{\bf e}\cdot{\bf m}({\bf r}_{\Gamma},\tau)]^2
\label{velo1} 
\end{eqnarray}
where ${\bf m}$ is the local normal to the interface,
$\kappa = \nabla\cdot{\bf m}$ is the local curvature, 
and $B^{\pm},S,I_0$ and $I_1$ are constants that
depend exclusively on the form of the conductivity
$\sigma$ and are given by Eqs. (\ref{B+-}),(\ref{S})
and (\ref{I}). The velocity depends on 
$\rho_1^{\pm}({\bf r},\tau)$ which satisfy
the {\it linear} partial differential equation
\begin{equation}
[~\nabla^2 - 2 B^{\pm}~{\bf e}\cdot\nabla~]~\rho_1^{\pm}({\bf r},\tau)=0
\label{eq:pde1} 
\end{equation} 
subjected to the boundary conditions 
\begin{eqnarray}
\label{eq:bc1}
\rho_1^+ ({\bf r}_{\Gamma},\tau)&=&
~-~2S~\kappa({\bf r}_{\Gamma},\tau)\;+\;
2 I_1~{\bf e}\cdot{\bf m}({\bf r}_{\Gamma},\tau) \\ 
\rho_1^- ({\bf r}_{\Gamma},\tau)&=&
~-~2S~\kappa({\bf r}_{\Gamma},\tau)\;+\;
2 (I_1-I_0)~{\bf e}\cdot{\bf m}({\bf r}_{\Gamma},\tau) 
\label{eq:bc2}
\end{eqnarray}
on the points ${\bf r}_{\Gamma}$ of the interface.

We are interested in the stability of a flat interface against
small perturbations in its profile. Let us consider an interface $\Gamma$ of the form 
\begin{equation}
\label{yps-gamma}
y_{\Gamma}(x,\tau) = h e^{i(kx-\omega \tau)} + V_0 \tau ~~~;~~~ k h << 1
\end{equation}
(see Fig. 1) separating two semi-infinite regions $\Omega^+$ 
(where $y>y_{\Gamma}(x,\tau)$) and
$\Omega^-$ (where $y<y_{\Gamma}(x,\tau)$). The direction ${\bf e}$ 
of the external electric field is arbitrary for the moment. To first order
in $k h$, we have
\begin{equation}
m_x = -i k h e^{i(k x - \omega \tau)} ~~~\mbox{and}~~~ m_y = 1
\end{equation}
for the $x$ and $y$ components of ${\bf m}$ respectively. The
local curvature $\kappa$ of the interface is given by
\begin{equation}
\kappa   = k^2 h  e^{i(k x - \omega \tau)} 
\end{equation}
while the normal velocity $V$ takes the form
\begin{equation}
V = 
-d y_s/d \tau =  -V_0 + i\omega h e^{i(k x - \omega \tau)}~~.
\label{eq:velo2}
\end{equation}

Eq. (\ref{eq:pde1}) has a solution $\rho_1^{\pm}({\bf r},\tau)$
of the form
\begin{equation}
\rho_1^{\pm}({\bf r},\tau) = f^{\pm}(\zeta) 
+ kh~e^{i(kx-\omega \tau)}g^{\pm}(\zeta) 
+ O(kh)^2
\label{eq:sol1}
\end{equation}
with $\zeta = y - he^{i(kx-\omega \tau)}$. 
The boundary conditions Eq. (\ref{eq:bc1}) and (\ref{eq:bc2})
 are satisfied for
\begin{eqnarray}
\label{fg}
f^+(\zeta) &=& 2 I_1 e_y - C^+~(1-e^{2B^+e_y\zeta}) \\
g^+(\zeta) &=& 
-[2S~ k + 2 ie_x I_1]e^{-\lambda^+\zeta} 
-2C^+\frac{B^+e_y}{k}~( e^{-\lambda^+\zeta} - e^{2B^+e_y\zeta})\\
f^-(\zeta) &=& 2(I_1-I_0) e_y - C^-~(1-e^{2B^-e_y\zeta}) \\
g^-(\zeta) &=& 
-[2S~ k + 2 ie_x(I_1-I_0)]e^{-\lambda^-\zeta} 
-2C^-\frac{B^-e_y}{k}~( e^{-\lambda^-\zeta} - e^{2B^-e_y\zeta})
\label{eq:sol2}
\end{eqnarray}
where $e_x$ and $e_y$ are the $x$ and $y$ components
of the unit vector ${\bf e}$ pointing in the direction of the
external field ${\bf E}\;$, and $\lambda^{\pm}$ are solutions of the 
following quadratic equation:
\begin{equation}
\label{lambda}
(\lambda^{\pm})^2 + 2 B^{\pm}e_y\lambda^{\pm} -k(k+2ie_x B^{\pm})=0 
\end{equation}
subjected to the constraint $Re(\lambda^{+}) > 0$ ($Re(\lambda^{-}) < 0$).
The new constants  $C^+$ and $C^-$ are determined by the boundary 
conditions at $z=\pm \infty$. We will discuss the effects of $C^{\pm}$
in the following section.

Inserting  $\rho_1^{\pm}({\bf r},\tau)$ given by Eqs. (\ref{eq:sol1})
- (\ref{eq:sol2}) in the equation for the normal velocity Eq (\ref{velo1}),
we can compare with Eq. (\ref{eq:velo2}) to obtain an expression for
the constant velocity $V_0$ 
\begin{eqnarray}
V_0 &=& B^+~(2I_1e_y-C^+)e_y - B^-~[~2(I_1-I_0)e_y-C^-]e_y \\
    &=& [B^+\rho_1^+(z=+\infty) - B^-\rho_1^-(z=-\infty)]~e_y
\end{eqnarray}
and for $\omega$
\begin{eqnarray}
i\omega &=& 
\frac{1}{2} k~ \frac{d}{dz}(g^+(z)-g^-(z))\vert_{z=0} 
+ k^2~[I_0 + 2S(B^+-B^-)]~e_y \nonumber \\
&+& 4 k~[I_0B^- + I_1(B^+-B^-)]~ie_xe_y
\end{eqnarray}

Assuming that $C^{\pm}$ are real, and writing 
$\lambda^{\pm}= \lambda_r^{\pm} + i\lambda_i^{\pm}$ and
$\omega = \omega_r + i\omega_i$, we get
\begin{eqnarray}
\omega_i &=& 
- k^2~[I_0+2S(B^+-B^-)]~e_y 
- k^2~S(\lambda_r^+-\lambda_r^-) \nonumber \\
&& + k~[I_1\lambda_i^+ - (I_1-I_0)\lambda_i^-]~e_x \nonumber \\
&& - C^+B^+ (\lambda_r^+ + 2B^+e_y) e_y
   + C^-B^- (\lambda_r^- + 2B^-e_y) e_y
\label{eq:stability}
\end{eqnarray}

A flat interface is unstable against small perturbation of
the form (\ref{yps-gamma}) if the external field, the conductivity and the
boundary conditions at infinity are such that $\omega_i > 0$.
Eq. (\ref{eq:stability}) is the main result of this paper. 

\subsection{The case of a third order conductivity}
In order to further study the properties of Eq. (\ref{eq:stability})
we assume in this section that the conductivity $\sigma(u)$ can
be written as a polynomial of the form
\begin{equation}
\label{sp sigma}
\sigma(u)= \sigma_0 + \sigma_1 u + \sigma_2 u^2 + \sigma_3 u^3
\label{eq:sigmapoly}
\;.\end{equation}
 Previous studies \cite{ymhj} of the stability
of a flat interface in a driven Cahn-Hilliard system of the form
Eqs (\ref{continuity})-(\ref{mu}) have  limited themselves to the case 
where $\sigma(u)= u(u_m-u)$. While we can reproduce their results, 
we will observe interesting new effects considering the more general 
form (\ref{eq:sigmapoly}) of $\sigma(u)\;$. 

The constants in Eq. (\ref{eq:stability}) can be expressed explicitly in terms 
of the system parameters. Using Eqs. (\ref{B+-}), (\ref{S}), (\ref{I}) and
(\ref{sigma1}) we find:
\begin{eqnarray}
\label{B+1}
B^+ &=& A  \;(\sigma_2  + 2\sigma_3 u_m )\\
\label{B-1}
B^- &=& - A\;  \;(\sigma_2  + \sigma_3 u_m) \\
\label{I0}
I_0 &=& - \sqrt{2}A \; (\sigma_2  + \frac{3}{2}\sigma_3 u_m) \\
\label{I1}
I_1 &=& - \sqrt{2}A\;(\frac{1}{2}\sigma_2  + \frac{5}{6}\sigma_3 u_m)\\
\label{S1}
S &=&\frac{\sqrt{2}}{12} 
\;\;\end{eqnarray} 
with  
\begin{eqnarray}
\label{A}
A=\epsilon^{-1} |{\bf E}| \xi \lambda^{-3/2} u_m^{-2}\;>\;0
\;\;.\end{eqnarray} 
All of the above constants are independent of $\sigma_0$ and $\sigma_1$. 
The reason for this is that the equations for the order parameter do not
depend on $\sigma_0$, and that a change of $\sigma_1$ amounts to a 
change of the reference frame only.

\subsubsection{External field parallel to interface}

In this case the direction of the field is such that
$e_x=\pm1$, $e_y=0$.
To lowest order in $k$ we get, using 
$\lambda_r^{\pm}\approx \pm k^{1/2}(|B^{\pm}|)^{1/2}$ and
$\lambda_i^{\pm}= \frac{k B^{\pm} e_x}{\lambda_r^{\pm}}$
\begin{equation}
\label{ey0}
\omega_i = k^{3/2}~[ I_1 \frac{B^+}{\sqrt{\vert B^+ \vert}} + 
(I_1-I_0) \frac{B^-}{\sqrt{\vert B^- \vert}} ]
\;.\end{equation}

Using the explicit expressions (\ref{B+1})-(\ref{A})  
one can show that $\omega_i < 0$, i.e. that the interface
is stable, for any choice of $\{\sigma_0,\sigma_1,\sigma_2,\sigma_3\}$.
This result does not depend on the sign of $e_x$, 
nor does is depend on the actual boundary
conditions away from the interface through the constants $C^{\pm}$.

\subsubsection{External field perpendicular to interface}

In this case we have $e_x=0$ and $e_y=+1$ if the external
field points into the high density region $\Omega^+$
or $e_x=0$ and $e_y=-1$ if it points towards the low density
region $\Omega^-$. We also have $\lambda_i^{\pm}=0$ from
Eq. (\ref{lambda}). 

Depending on the  signs of $B^+e_y$ and $B^-e_y$ 
we get different conditions on the possible values of
the constants $C^{\pm}$. From Eqs (\ref{fg}) we see that if
$B^+e_y >0$ ($B^-e_y < 0$) we must take $C^+=0$ ($C^-=0$)
in order for $\rho_1^+$ ($\rho_1^-$) to be finite as
$z\rightarrow \infty$ ($z\rightarrow -\infty$). This means
that the system does not support a single interface in this case
unless there is some current of particles at infinity.
(see also the discussion of the one-dimensional
model in Ref. \cite{eb})

If $B^+e_y <0$ ($B^-e_y > 0$) then there are no such
constraints on $C^+$ ($C^-$) and we chose
the special cases $C^+= 2I_1e_y$ ($C^-= 2(I_1-I_0)e_y$)
which makes $\rho_1^+$ ($\rho_1^-$) vanish at infinity.

We are left with four cases to study for $\omega_i$: 

\begin{itemize}
\item {\bf CASE A}: $B^+ e_y > 0$ and $B^- e_y <0$. 
If we assume that $k << B^{\pm}$, then 
$\lambda_r^{\pm}\approx \frac{1}{2} \frac{k^2e_y}{B^{\pm}}$
and we get, to lowest order,
\begin{eqnarray}
\omega_i &=& -k^2e_y~[I_0+2S(B^+-B^-)]
\end{eqnarray}

\item {\bf CASE B}: $B^+ e_y <0$ and $B^- e_y >0$.
To lowest order we write 
$\lambda_r^{\pm} = -2 B^{\pm}e_y~(1+\frac{k^2}{4(B^{\pm})^2}
-\frac{k^4}{16(B^{\pm})^4})$, and get
\begin{eqnarray}
\omega_i &=& \frac{-k^4 e_y}{4(B^+)^2(B^-)^2}
~[I_1(B^-)^2-(I_1-I_0)(B^+)^2 +2S(B^+-B^-)B^+B^- ]
\end{eqnarray}

\item {\bf CASE C}: $B^+ e_y <0$ and $B^- e_y <0$. 
With $\lambda_r^+ = -2B^+e_y(1+\frac{k^2}{4(B^+)^2})$, 
$\lambda_r^- = -\frac{k^2e_y}{2B^-}$, we get
\begin{equation}
\omega_i = -k^2e_y~(I_0 -I_1 -2S~B^-)
\end{equation}

\item {\bf CASE D}: $B^+ e_y >0$ and $B^- e_y >0$.
This is essentially the reverse of the previous case.
$\lambda_r^+ = \frac{k^2e_y}{B^+}$, 
$\lambda_r^- = -2B^-e_y(1+\frac{k^2}{4(B^-)^2})$, and get
\begin{equation}
\omega_i = -k^2e_y(I_1 +2S~B^+)
\end{equation}
\end{itemize}

Inserting the explicit expressions (\ref{B+1})-(\ref{A}) for 
$B^{\pm}\;,\;I_0\;,\;I_1$ and $S$ into above equations for
$\omega_i$ one comes to the following conclusion:
In all four cases the interface is unstable if
\begin{equation}
(2\sigma_2  +  3\sigma_3 u_m) e_y > 0 
\label{condi}
\;\;\end{equation}
and stable otherwise. Condition (\ref{condi}) is the main result of this
section.  

It is natural to assume the conductivity to be
positive in the whole interval $[0,u_m]\;$. 
Assuming furthermore that $\sigma(u)$ is symmetric around $u_m$ like the chemical
potential and vanishes for $u=0\;$, we are lead to the standard
expression  $\sigma(u)=u(u_m-u)$  \cite{ymhj}. There 
$\sigma_2<0$ and $\sigma_3=0\;$, the interface is  unstable  if $e_y=-1$,
i.e. the case where the external field points away from the
high density region $\Omega^+$ \cite{ymhj}. Although the assumptions of symmetry
and vanishing $\sigma(0)$ are natural for certain Ising-like \cite{ppd}-\cite{bray}, \cite{ymhj}, \cite{eb}, \cite{allz} or particle-hopping models \cite{gl},
one can think of more general situations (e.g. the microscopic
model in \cite{ws}). Then,  
the parameters  $\{\sigma_0,\sigma_1,\sigma_2,\sigma_3\}$ can be
such that $(2\sigma_2  +  3\sigma_3 u_m)>0 \;$. In this case the interface
is unstable for $e_y=+1\;$, i.e. if the external field points into the
high density region $\Omega^+$.

\section{Summary}

We have applied a matched asymptotic expansion
to a system of equations describing the dynamics of 
phase segregation in the presence of an external field.
The influence of the field on a region of local density
$u$ is given by a conductivity $\sigma(u)$. We derived
 equations for the dynamics of an interface separating
two regions in different phases 
(see Eqs. (\ref{rho_1}), (\ref{u1plus}), (\ref{u1minus})
and (\ref{vsol}) ) and studied the stability 
of a flat interface
against small perturbations. We found a general condition (see Eq. (\ref{eq:stability}) )
 for the stability of such an interface and discussed in more detail
the  case where $\sigma(u)$ can be written
as a third order polynomial in $u\;$.
In this case  the interface is always
stable if the field is parallel to it. 
However, if the field is perpendicular to the interface, 
 the interface can be either stable or unstable depending on
the conductivity and on the direction of the field (see Eq. (\ref{condi}) ).

\vskip 1cm
{\bf \Large Acknowledgements}\\
We thank  B. Schmittmann  and R. K. P. Zia  for stimulating discussions. 
S. S. gratefully acknowledges financial support by the 
Deutsche Forschungsgemeinschaft. C. A. L. has been supported by the National 
Science Foundation through DMR-9419393.

\vspace{5cm}

{\bf \Large Caption to the figure}
\vskip 12pt
\noindent
{\em Fig.1:} Schematic picture of the approximately flat interface $\Gamma$
defined by Eq. (\ref{yps-gamma}). The interface separates the semi-infinite
regions $\Omega^+$ and $\Omega^-\;$, in which $\rho \approx 1$ and 
$\rho \approx 0\;$, respectively. In the figure ${\bf m}$ denotes
the unit normal vector on $\Gamma$ and $h$ the amplitude of the perturbation
around a flat interface.  The unit normal vector of the external
field is denoted by $\hat {\bf e}$.


\begin{thebibliography}{99}
\bibitem{ch}
      J. W. Cahn and J. E. Hilliard, J. Chem. Phys. {\bf 28}, 258 (1958) 
\bibitem{gss}
      J. D. Gunton, M. San Miguel and P. S. Shani,
      in {\em Phase Transitions and Critical Phenomena}, Vol 8,
      C. Domb and J. L. Lebowitz, eds., Academic Press, New York 1983
\bibitem{langer}
      J. S. Langer, {\em An Introduction to the Kinetics of first-order
       Phase Transitions}, in {\em Solids far from Equilibrium},
      C. G\`odreche, ed.,
       Cambridge University Press (1991)    
\bibitem{koj}
       K. Kitahara, Y. Oono and D. Jasnow,
       Mod. Phys. Lett. B {\bf 2}, 765 (1988) 
\bibitem{lbm}
       J. S. Langer, N. Bar-On and H. D. Miller, 
       Phys. Rev. A {\bf 11}, 1417 (1975)
\bibitem{ppd}
      S. Puri, N. Parekh and S. Dattagupata, J. Stat. Phys. 
      {\bf 75}, 839 (1994) 
\bibitem{lhs}
       A. M. Lacasta, A. Hern\'andez-Machado and J. M. Sancho ,
       Phys. Rev. B {\bf 48}, 9418 (1993)  
\bibitem{bray}
      A. J. Bray, Adv. Phys. {\bf 43}, 357 (1994)
\bibitem{pego}
      R. L. Pego, Proc. R. Soc. Lond. {\bf A 422}, 261 (1989)  
\bibitem{ymhj}
      C. Yeung, J. L. Mozos, A. Hern\'anez-Machado and D. Jasnow,
         J. Stat. Phys. {\bf 70}, 1149 (1993)
\bibitem{gl}
      G. Giacomin and J. L. Lebowitz, Phys. Rev. Lett. {\bf 76}, 1094 (1996)
\bibitem{eb}
      C. L. Emmot and A. J. Bray, Phys. Rev. E {\bf 54}, 4568 (1996)
\bibitem{allz}
      F. J. Alexander, C. A. Laberge, J. L. Lebowitz and R. K. P. Zia
       J. Stat. Phys. {\bf 82}, 1133 (1996)
\bibitem{sz}
	B. Schmittmann and R. P. K. Zia, {\it Statistical Mechanics of
		Driven Diffusive Systems}, in {\it Phase Transitions
		and Critical Phenomena}, vol. 17, ed. by. Domb and 
		Lebowitz, Academic Press, U.K. (1995)
\bibitem{ws}
      L. K. Wickham and J. P. Sethna,
      Phys. Rev. B {\bf 51}, 15017 (1995)


\end{thebibliography}
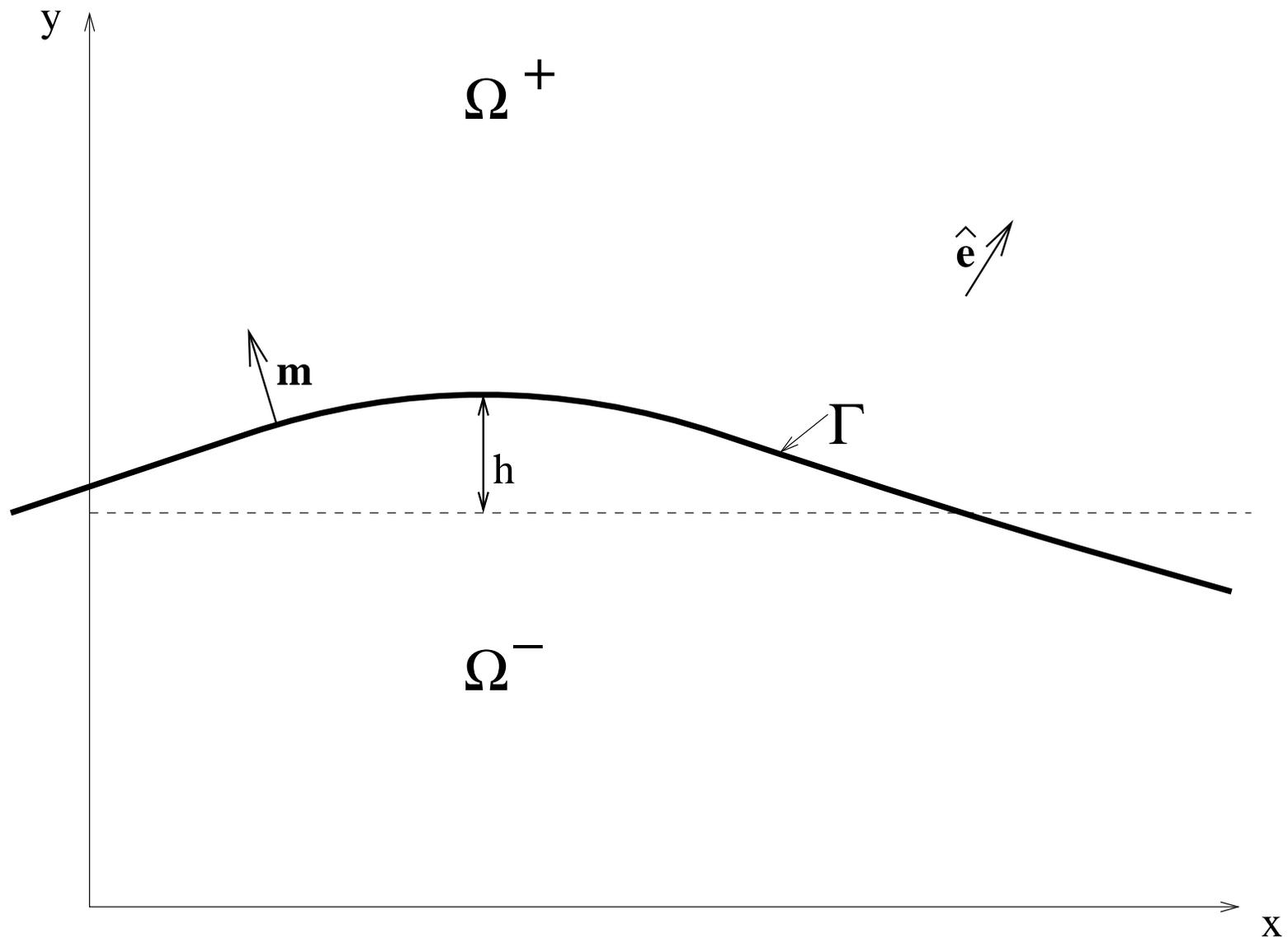
\end{document}